\documentclass[prl,twocolumn,showpacs,preprintnumbers,amsmath,amssymb,superscriptaddress,nofootinbib]{revtex4}

\usepackage{graphics}
\usepackage{graphicx}
\usepackage{epsfig}
\usepackage{epsf}
\usepackage{bm}
\usepackage{subfigure}
\usepackage{amsmath}
\usepackage{color}

\usepackage{amssymb}

\begin{document}




\title{Combining Harmonic Generation and Laser Chirping to Achieve \\ High Spectral 
Density in Compton Sources}


\author{Bal{\v s}a Terzi{\'c}}
\thanks{Email: bterzic@odu.edu} 
\affiliation{Department of Physics, Old Dominion University, Norfolk, Virginia 23529, USA}
\affiliation{Center for Accelerator Science, Old Dominion University, Norfolk,
Virginia 23529, USA}

\author{Cody Reeves}
\affiliation{Jefferson Lab, Newport News, Virginia 23606, USA}

\author{Geoffrey A.~Krafft}
\affiliation{Department of Physics, Old Dominion University, Norfolk, Virginia 23529, USA}
\affiliation{Center for Accelerator Science, Old Dominion University, Norfolk,
Virginia 23529, USA}
\affiliation{Jefferson Lab, Newport News, Virginia 23606, USA}


\begin{abstract}
Recently various laser-chirping schemes have been investigated with the goal of 
reducing or eliminating ponderomotive line broadening in Compton or Thomson 
scattering occurring at high laser intensities. As a next level of detail in the spectrum 
calculations, we have calculated the line smoothing and broadening expected due to 
incident beam energy spread within a one-dimensional plane wave model for the 
incident laser pulse, both for compensated (chirped) and unchirped cases. The scattered  
compensated distributions are treatable analytically within three models for the 
envelope of the incident laser pulses: Gaussian, Lorentzian, or hyperbolic secant.
We use the new results to demonstrate that the laser chirping in Compton sources
at high laser intensities: (i) enables the use of higher order harmonics, thereby reducing 
the required electron beam energies; and (ii) increases the photon yield in a small
frequency band beyond that possible with the fundamental without chirping. This 
combination of chirping and higher harmonics can lead to substantial savings in the 
design, construction and operational costs of the new Compton sources.
This is of particular importance to the the widely popular laser-plasma accelerator
based Compton sources, as the improvement in their beam quality enters the regime 
where chirping is most effective.
\end{abstract}


\maketitle

Compton or Thomson sources are increasingly being considered as potential 
sources of high-energy photons \cite{krafftreview, ruth}. A principal attraction of 
such sources is the narrow bandwidth generated in the output radiation. 
As a narrow bandwidth is desired, it is important to understand the sources of 
line width in the scattered radiation and to eliminate them to the extent possible. 
Recently it has been shown by a Jefferson Lab group \cite{terzic} (hereafter TDHK2014)
and others \cite{gheb, seipt2015} that the linewidth can be preserved 
against ponderomotive line broadening \cite{krafft} (hereafter K2004)
even at high laser intensities by proper chirping of the laser pulse. 

It has been understood for many years in the FEL community \cite{m70,c81,bm89}
that harmonic generation can provide a path to a given radiation wavelength
with a smaller electron energy than is needed for fundamental emission,
or that harmonic generation provides access to shorter wavelengths for
a given electron beam energy. This idea, as applied to Compton sources,
suffers from the fact that whenever the field strength is low enough that 
ponderomotive broadening is negligible, emission into the harmonics falls off rapidly 
with harmonic number. On the other hand, if the field strength is large enough for
substantial harmonic emission, ponderomotive broadening limits the
spectral density of the emission, increasingly at high harmonic number.
The main motivation for our present work is to point out how
important it is that the new chirping prescriptions remove
ponderomotive broadening from the harmonics, allowing high emitted spectral
density in the harmonics at high field strength.

In this work we have calculated spectra for both the compensated and uncompensated 
cases in regimes where the normalized vector potential is of order 1, i.e.~where significant 
ponderomotive broadening is expected. We then use the 
new results to demonstrate that the laser chirping in Compton sources enables
them to retain the narrowband radiation and increase the photon yield in all harmonics 
at high laser intensities. The increase in the photon yield enables the efficient use of 
higher-order harmonics, which greatly reduces the strain on the electron beam source. 
Combining laser chirping and the higher harmonics is particularly beneficial for 
the Compton sources based on the laser-plasma accelerators, as their improved
beam energy spread enters the range in which the chirping is most effective.

Ponderomotive line broadening is due to the variable red-shifting of the emitted radiation
because the longitudinal velocity of the electrons changes within the incident laser pulse.  
Compensating the local value of the frequency in the incident laser pulse against the 
ponderomotive longitudinal velocity change to yield constant emitted frequency through 
the (relativistic) Doppler shift leads to analytic expressions for the proper frequency 
modulation (FM) to achieve {\it perfect} compensation \cite{terzic}. It is perfect because
the bandwidth of the backscattered radiation is reduced to that of the incoming laser pulse, 
thereby reaching the physical limit.  
%

Here, as in TDHK2014, calculations are completed using the formalism
developed in K2004 for the 
far-field spectral distribution of photons Thomson-scattered by a single
electron. The incident laser pulse is described by a plane wave. The 
treatment is fully relativistic and includes the classical electron motion without 
approximation. We assume a linearly polarized incident plane wave described
by a single component for the normalized vector potential 
${\tilde{A}(\xi})=eA(\xi)/mc=a(\xi)\cos(2\pi\xi f(\xi)/\lambda)$ where 
$a(\xi)$ describes the envelope of the oscillation, $\xi=z+ct$ is the 
coordinate along the laser pulse, $f(\xi)$ specifies the laser FM, and $\lambda$ is a 
normalizing wavelength. 

Expressions for backscattered radiation spectra in high-intensity Compton 
sources emitted by a beam with an energy spread are derived from the 
equations for scattering off a single electron. K2004 derives 
expressions for the constant-frequency laser pulses, while TDHK2014
provides the spectra for FM laser pulses.

Both derivations for the backscattered radiation spectrum 
$(d^2 E/d\omega d\Omega)_{\rm beam}$ start with 
\begin{equation} \label{d2I}
\left({{d^2 E(\omega)}\over{d\omega d\Omega}}\right)_{\rm beam} =  \int_1^{\infty} N(\gamma) 
{{d^2 E(\gamma, \omega)}\over{d\omega d\Omega}} d\gamma, 
\end{equation}
where $\omega$ is the frequency of the scattered radiation, $\Omega$ is the solid angle
of the radiation, $\gamma$ is the relativistic factor, $N(\gamma)$ is the beam's energy 
distribution and $d^2 E/d\omega d\Omega$ the radiation of a single electron.
The single-electron scattering spectrum can be expressed as in TDHK2014
\begin{equation} \label{dI3} 
{{d^2 E(\gamma, \omega)}\over{d\omega d\Omega}} = 
\left({{d^2 E(\gamma)}\over{d\omega d\Omega}}\right)_n 
\left({{\omega}\over{\omega_0(\gamma)}}\right)^2 
\left|{{(1+\beta)\gamma D_x}\over{\lambda}}\right|^2, 
\end{equation}
where $\omega_0 (\gamma) = (1+\beta)^2 \gamma^2 2 \pi c/\lambda$ is the normalizing
frequency, $(d^2 E (\gamma)/d\omega d\Omega)_n=(1+\beta)^2\gamma r_eE_{\rm beam}/c$ 
is the normalization factor and $c$ the speed of light. We recast the frequency 
content form $D_x$ from Eq.~(7) of TDHK2014
\begin{eqnarray} \label{eq:Ds}
{\tilde D}_x 
& \equiv & 
{{(1+\beta)\gamma D_x \left(\gamma, \omega\right)}\over{\lambda}} \\
& = &   
{1\over 2} {\int^\infty_{-\infty}a(\xi)
\exp\left[-2\pi i 
\left(\xi f(\xi) + {{\omega}\over{\omega_0(\gamma)}} {\tilde Z}(\xi) \right)\right]}  d\xi,
\nonumber
\end{eqnarray}
where ${\tilde Z}(\xi) = \xi + \int_{-\infty}^\xi\tilde{A}^2(\xi')d\xi'$ and the FM function
\begin{equation} \label{eq:exact}
f(\xi)={1\over{1+a(0)^2/2}}
\left(1+{\int_0^\xi a(\xi')^2d\xi'\over2\xi}\right).
\end{equation}
Therefore, the beam energy only scales the spectrum in both frequency
and amplitude: 
\begin{equation} \label{dI3b} 
{{d^2 E(\gamma, \omega)}\over{d\omega d\Omega}} = 
\left({{d^2 E(\gamma)}\over{d\omega d\Omega}}\right)_n 
\left({{\omega}\over{\omega_0(\gamma)}}\right)^2 
\left|{\tilde D}_x \left({\omega\over{\omega_0(\gamma)}}\right) \right|^2. 
\end{equation}
This means that for each experiment, it suffices to compute only 
one scale-free single-electron spectrum, ${\tilde D_x}(\omega/\omega_0)$, 
as in K2004 and TDHK2014. 
The computation of Eq.~(\ref{d2I}) is then reduced to integration of that 
scale-free spectrum, properly shifted and scaled. This treatment applies to both constant 
frequency scattering of K2004 and FM scattering in TDHK2014, and to all harmonics.
%

Figure 2(b) in TDHK2014 demonstrates the perfect agreement between the 
frequency bandwidth of the incoming laser pulse and the bandwidth of the 
backscattered radiation after the FM, and the near-agreement in their maximum
amplitudes. Therefore, to first order, one can approximate the first harmonic of 
the backscattered radiation spectrum for a FM laser pulse by its Fourier 
transform:
\begin{equation} \label{Dxi1}
{\tilde D}_x  \left({\omega\over{\omega_0(\gamma)}}\right) = {1\over 2} {\cal F} \left\{ a(\xi) \right\} \left({{{\omega\over{\omega_0(\gamma)}}-{\tilde w}_f}\over{{\tilde w}_f}}\right).
\end{equation}
The peak width is scaled by ${\tilde w}_f = 1/(1+a_0^2/2)$ due to the first-order 
expansion of ${\tilde Z}(\xi)$ around $\xi=0$ in Eq.~(\ref{eq:Ds}).
For the three laser pulse shapes considered here: 
Gaussian $a_G(\xi) = a_0 \exp\left[-\xi^2/(2(\sigma\lambda)^2)\right]$,
Lorentzian $a_L(\xi) = a_0 \sigma/((\xi/\lambda)^2 + \sigma)$ and hyperbolic secant 
$a_S(\xi) = a_0 {\rm sech}~(\sigma\xi/\lambda)$, they are:
\begin{subequations} \label{Ds}
\begin{eqnarray} 
{\tilde D}_{x, G} \left({\omega\over{\omega_0}}\right) & = & a_0 \sigma \sqrt{\pi\over{2}}
\exp\left(-{{ \left({{\omega}\over{\omega_0}} - {\tilde w}_f\right)^2}\over{2 {\bar \Sigma}^2}}\right), \label{DxG}  \\
{\tilde D}_{x, L} \left({\omega\over{\omega_0}}\right) & = & {{a_0 \pi \sigma} \over {4}}
\exp\left(-{{\left|  {{\omega}\over{\omega_0}} - {\tilde w}_f \right|}\over{2 {\bar \Sigma}}} \right), \label{DxL}  \\
{\tilde D}_{x, S} \left({\omega\over{\omega_0}}\right) & = & {{a_0 \pi} \over {2\sigma}} 
{\rm sech} \left({{\pi^2}\over{{\bar \sigma}}}\left( {{\omega}\over{\omega_0}} - {\tilde w}_f \right)\right), \label{DxS} 
\end{eqnarray}
\end{subequations}
where $\Sigma = 1/2\pi \sigma$, ${\bar \Sigma} = \Sigma {\tilde w}_f$
and ${\bar \sigma} = \sigma {\tilde w}_f$. These are achieved by FM: 
\begin{subequations} \label{fs}
\begin{eqnarray} 
f_G({\bar \xi}) & = & {\tilde w}_f
\left(1+{{\sqrt{\pi}\sigma a_0^2}\over{4{\bar \xi}}} {\rm erf}({\bar \xi}/\sigma)\right), \label{fG}  \\
f_L({\bar \xi}) & = & {\tilde w}_f 
\left(1+{{\sqrt{\sigma} a_0^2}\over{4\sqrt{2}{\bar \xi}}} \left({{\sqrt{2\sigma}{\bar \xi}}\over{\sigma+2{\bar \xi}^2}} + 
\tan^{-1}{{\sqrt{2}{\bar \xi}}\over{\sqrt{\sigma}}}\right)\right), \label{fL} \\
f_S({\bar \xi}) & = & {\tilde w}_f 
\left(1+{{a_0^2}\over{2{\bar \xi}\sigma}} {\rm tanh}(\sigma{\bar \xi})\right), \label{fS}   
\end{eqnarray}
\end{subequations}
with ${\bar \xi} = {\xi}/\lambda$.

\begin{figure}
\begin{center}
\subfigure{
\includegraphics[width=3in]{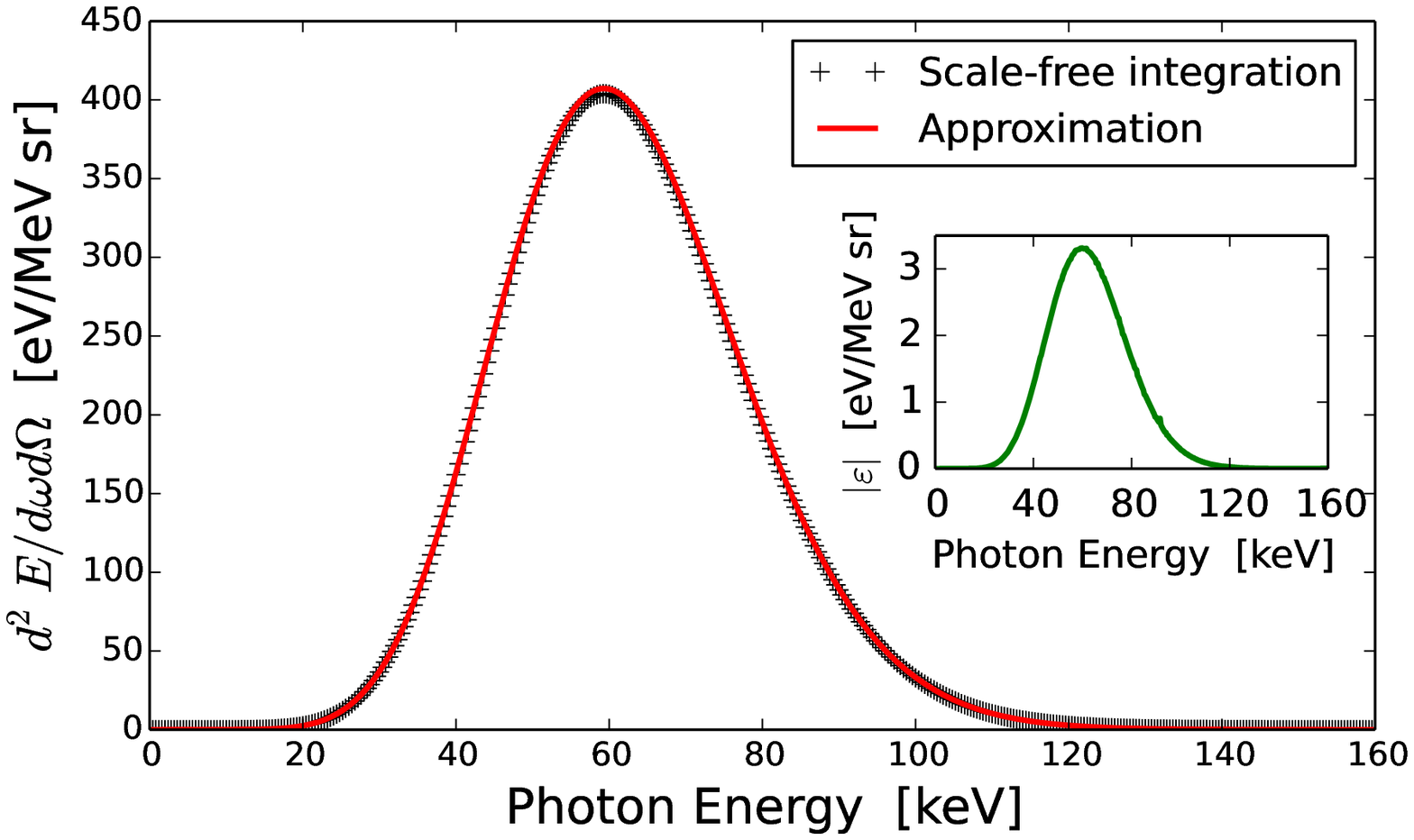} \label{fig:appx_sf}}
\vskip-8pt
\subfigure{
\includegraphics[width=3in]{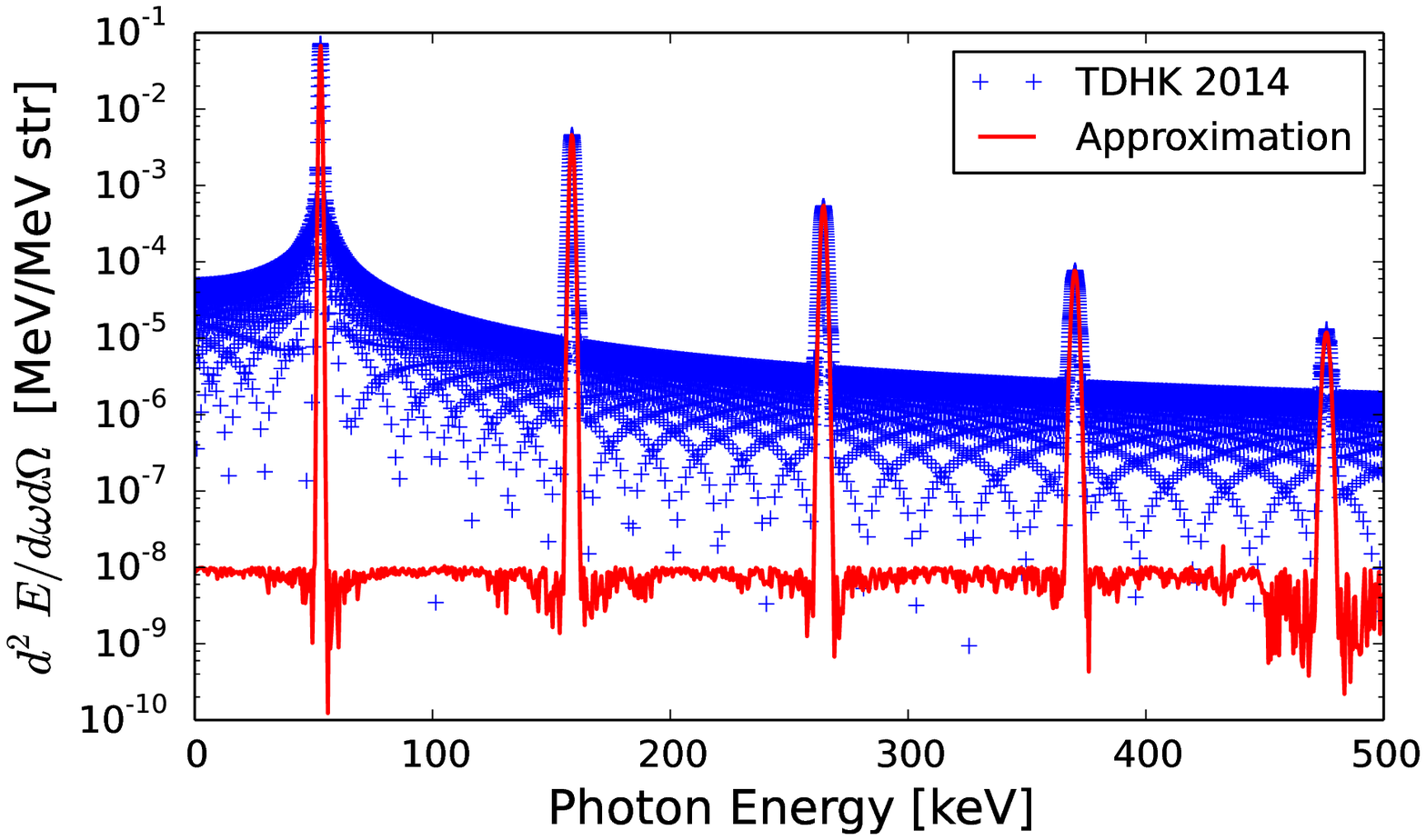} \label{fig:appx_hoh}}
\vskip-5pt
\caption{\small{Spectra from scattering a FM Gaussian laser
pulse with $\lambda = 800~{\rm nm}$, $a_0 = 0.587$ off a Gaussian electron beam 
with $Q=100~{\rm pC}$ and $E_0 = 51.1~{\rm MeV}$. 
(a) Integration of a single scale-free spectrum ${\tilde D}_x$ and an electron 
beam with a 34\% energy spread versus the approximation from Eq.~(\ref{d2I_FM_G}) (red line). 
The inset shows the absolute difference $\varepsilon$ between the two results.
(b) Single-electron scattering approximation in Eq.~(\ref{eq:Ds3}) against the exact 
solution of TDHK2014.
}}
\label{fig:appx}
\end{center}
\end{figure}
%
\begin{figure}
\begin{center}
\includegraphics[width=3in]{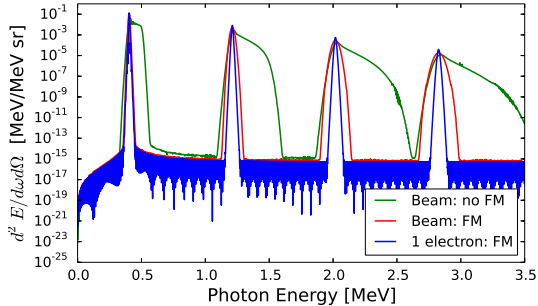}
\vskip-5pt
\caption{\small{Backscattered radiation of a Gaussian 
laser pulse with $\lambda$ = 1~$\mu$m, 
$a_{0}$ = 0.707 off a Gaussian electron beam with Q = 100~pC, 
$E_0$ = 163~MeV, and 1\% FWHM energy spread: FM (red), non-FM
(green), and the single electron FM case (normalized to FM)  (blue). 
Lorentzian and hyperbolic secant behave similarly.
}} 
\label{fig:spec}
\end{center}
\end{figure}

Obtaining an analytic approximation for the shape of the higher-order harmonics in
the FM spectrum requires a derivation similar to that in \cite{Brau}. After 
substituting Eq.~(\ref{eq:exact}) into Eq.~(\ref{eq:Ds}), integrating by parts, using
$\exp(i\alpha \sin \theta) = \sum_{n=-\infty}^{\infty} J_n(\alpha) \exp(in\theta)$, where
$J_n$ is the Bessel function of the $n$th order, and expanding around the stationary
phase point, we find the contribution of the $n$th harmonic to the scale-free spectrum 
${\tilde D}_x = \sum_n {\tilde D}_x^n$ 
\begin{equation} \label{eq:Ds2}
{\tilde D}_x^n =
 \int^\infty_{0}a(\xi)
K_n \left(g(\xi)\right)
\cos\left(
2\pi Z(\xi){\bar \omega}_n
\right) d\xi,
\end{equation}
with $K_n(\alpha) = (-1)^n\left[J_n(\alpha) - J_{n-1}(\alpha)\right]$, 
$g(\xi) = a^2(\xi)\left( n - {1/2}\right)/(2\left(1+{1/2}a^2(\xi)\right)$,
${\bar \omega}_n = \omega/\omega_0 - 2(n-1/2){\tilde w}_f$ and 
$Z(\xi) = \xi + (1/2) \int_0^{\xi} a^2(\xi') d\xi'$. Examining 
$K_n(g(\xi))$, an implicit function of $\xi$, reveals that the explicit function 
${\tilde K}_n(\xi)$ is well-approximated by a Gaussian 
${\tilde K}_n(\xi) \approx A_n \exp\left(-{{\xi^2}/{2\Sigma_n}}\right)$ for $n>1$ and
a shifted Gaussian 
${\tilde K}_1(\xi) \approx 1+ A_1 \exp\left(-{{\xi^2}/{2\Sigma_1}}\right)$ for $n=1$,
where $A_1 = K_1(g(0)) - 1$, $A_n = K_n(g(0))$, 
$\Sigma_1 = \sqrt{-B_1^2/2\log((B_1-1)/A_1)}$, 
$\Sigma_n = \sqrt{-B_n^2/2\log(B_n/A_n)}$
$B_n = K_n(g(\sigma))$. Keeping only the linear term in $Z(\xi)$, we obtain 
analytic approximation for all harmonics:
\begin{subequations} \label{eq:Ds3}
\begin{eqnarray}
{\tilde D}^1_x & = & {1\over 2} {\cal F} \left\{ a(\xi) \left(1 + A_1\exp\left[-{{\xi^2}\over{2\Sigma_n^2}}\right] \right) \right\} 
\left({{{\bar \omega}_1}\over{{\tilde w}_f}}\right), \\
{\tilde D}^n_x & = & {1\over 2} {\cal F} \left\{ a(\xi) A_n \exp\left[-{{\xi^2}\over{2\Sigma_n^2}}\right] \right\} 
\left({{{\bar \omega}_n}\over{{\tilde w}_f}}\right). 
\end{eqnarray}
\end{subequations}
Neglecting a correction due to the small parameter $A_1$, the 
result for the leading-order harmonic in Eq.~(\ref{Dxi1}) is recovered. 
An excellent agreement between this approximation and the exact scale-free solution 
is shown in Fig.~\ref{fig:appx_sf}.
For the higher-order harmonics, only Gaussian pulse yields an analytic approximation,
shown in Fig.~\ref{fig:appx_hoh}. The agreement is nearly perfect. 
The Eq.~(\ref{eq:Ds3}) suggests that only for the Gaussian laser pulse is the shape of all
the harmonics the same. The higher-order harmonics of 
the Lorentzian and hyperbolic secant pulse are different from their respective first harmonics. 

The maximum amplitude of the harmonics is computed from Eq.~(\ref{eq:Ds3}) by
substituting $\int_{-\infty}^{\infty} d\xi$ for ${\cal F} \left\{ \right\}$:
\begin{subequations} \label{Ds1n}
\begin{eqnarray}
\left| {\tilde D}^{1,n}_x \right|_{{\rm max}, G} & = & 
{a_0 {\sqrt{\pi\over 2}}} \left(\sigma + A_1 s^1_{G},  A_n s^n_G \right), \\
\left| {\tilde D}^{1,n}_x \right|_{{\rm max}, L} & = & a_0 \left(\pi \sqrt{\sigma} + A_1 s^1_{L}, 
A_n s^n_{L} \right),
\end{eqnarray}
\end{subequations}
where $s^n_{G} = \sigma \Sigma_n/\sqrt{\Sigma_n^2 + \sigma^2}$, 
$s^n_{L} = (\pi/2\sqrt{\sigma}) \exp\left(\beta_n^2\right) \left[1 - {\rm erf} (\beta_n)\right]$ 
and $\beta_n = \sqrt{\sigma/2\Sigma_n^2}$. 
The width of the harmonics for the Gaussian pulse are 
%
\begin{equation}
W^1_{G} \approx {{\tilde w}_f\over{2\pi \sigma}}, \hskip10pt
W^n_{G} = {{\tilde w}_f\over{2\pi s_{G}^n}}.
\end{equation}
%
For FWHM in eV, these are multiplied by 2.35$\hbar \omega_0$.

For the Gaussian electron beam distribution and the case with FM, substituting Eq.~(\ref{Ds})
into Eq.~(\ref{dI3b}) and Eq.~(\ref{d2I}) leads to analytic solutions after invoking 
a well-justified approximation that the narrowband backscattered radiation 
off a single electron ${{d^2 E(\gamma, \omega)}/{d\omega d\Omega}}$ is highly peaked
around ${\bar \gamma} (\omega) = \sqrt{\lambda \omega/2\pi c {\tilde w}_f}$. 
Then the single-electron spectra are well-approximated by a Gaussian, which 
exactly integrates into analytic expressions for the leading-order harmonic:
\begin{widetext}
\begin{subequations} \label{d2I_analytic}
\begin{eqnarray}
\left({{d^2 E(\omega)}\over{d\omega d\Omega}}\right)_{\rm beam, FM, G} & = &
{{Q e r_e a_0^2 \sigma^2 \lambda {\tilde w}_f}\over{4 q_e c^2}}
{{S_G(\omega) \omega}\over{\sqrt{{\tilde \sigma}_E^2 + S_G(\omega)^2}}}
\exp\left[-{{(\gamma_0-{\bar \gamma}(\omega))^2}\over{2({\tilde \sigma}_E^2 + S_G^2(\omega))}}\right],
\label{d2I_FM_G} \\
\left({{d^2 E(\omega)}\over{d\omega d\Omega}}\right)_{\rm beam, FM, L} & = &
{{Q e r_e a_0^2 \sigma^2 \lambda {\tilde w}_f \pi^3}\over{32\sqrt{2} q_e c^2 {\tilde \sigma}_E}}
\left[
r_{+} \left(1 - {\rm erf} (s_{+})\right) - r_{-} {\rm erfi} (s_{-}) \right],
\label{d2I_FM_L} \\
\left({{d^2 E(\omega)}\over{d\omega d\Omega}}\right)_{\rm beam, FM, S} & = &
{{Q e r_e a_0^2 \lambda {\tilde w}_f \pi k}\over{8 q_e c^2 \sigma^2}}
{{S_S(\omega) \omega}\over{\sqrt{{\tilde \sigma}_E^2 + S_S(\omega)^2}}}
\exp\left[-{{(\gamma_0-{\bar \gamma}(\omega))^2}\over{2({\tilde \sigma}_E^2 + S_S^2(\omega))}}\right],
\label{d2I_FM_S} 
\end{eqnarray}
\end{subequations}
\end{widetext}
where 
$S_G(\omega) = {{{\bar \gamma} (\omega) \Sigma}/({2\sqrt{2}}} {\tilde w}_f)$,
$S_L(\omega) = {{{\bar \gamma} (\omega)}/({\sqrt{2\pi {\tilde w}_f}}} (2\sigma)^{1/4})$,
$S_S(\omega) = {{{\bar \gamma} (\omega) \sigma}/({2\sqrt{2} \pi^2 {\tilde w}_f)}}$, 
$k=1.12841$ is the normalization factor, erfi the complex error function
${\rm erfi}(x) = i {\rm erf}(ix)$ and
\begin{equation}
r_{\pm} = {{\exp\left[{{-\gamma_0^2}\over{2({\tilde \sigma}_E^2 \pm S_L^2)}} \mp 2\sqrt{2\sigma} \pi {\tilde w}_f \right]}\over{\sqrt{{\tilde \sigma}_E^2 \pm S_L^2}}},  \nonumber
s_{\pm} = {{{\bar \gamma}{\tilde \sigma}_E^2 \pm S_L^2({\bar \gamma}-{\gamma}_0)}\over{\sqrt{2}{\tilde \sigma}_E S_L \sqrt{{\tilde \sigma}_E^2\pm S_L^2}}}. \nonumber
\end{equation}
A single-electron scattering is recovered when
${\tilde \sigma}_E \to 0$ and $Q=q_e$ are substituted. 
Figure \ref{fig:appx_sf} shows an excellent agreement between the analytic 
approximation Eq.~(\ref{d2I_FM_G}) and the numerical integration
of Eq.~(\ref{d2I}) with a single scale-free spectrum ${\tilde D}_x$ given in Eq.~(\ref{eq:Ds}).
The accuracy for other laser pulse shapes is similarly high. 
Clearly, one can confidently use either of these methods for the FM laser pulses---direct
scale-free numerical integration or the analytic approximation.
However, for the constant-frequency laser pulses for which the single-electron scattering
${{d^2 E(\gamma, \omega)}/{d\omega d\Omega}}$ can only be numerically computed as 
in K2004, only the scale-free integration works. 

Figure \ref{fig:spec} shows the spectrum of the backscattered radiation from a electron 
beam with a 1\% energy spread with and without FM, and from a single-electron 
scattering. Clearly, the FM is still quite effective in
restoring the narrowband spectrum for all harmonics simultaneously. 

\begin{figure} 
\includegraphics[width=3in]{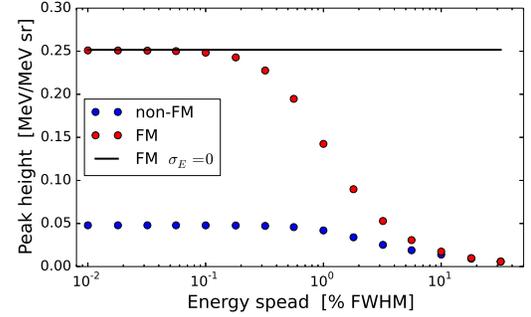} \label{fig:high}
\caption{\small{Maximum amplitude as a function of the energy spread for a
Gaussian laser pulse with $\lambda = 1~\mu {\rm m}$, $a_0 = 0.707$ off a 
Gaussian electron beam with $Q = 100~{\rm pC}$, $E_0 = 163~{\rm MeV}$ 
without FM (blue dots) and with FM (red dots), along with the theoretical limit 
for the FM monochromatic ($\sigma_E=0$) beam from Eq.~(\ref{d2I_FM_G}) 
(solid black line), computed by scale-free integration.
The results for the Lorentzian and hyperbolic secant are virtually identical.}}
\label{fig:hw}
\end{figure}

The bandwidth of the backscattered radiation---for both the FM and constant-frequency
laser pulse---is affected by: (1) the intrinsic bandwidth of the single-electron 
scattering (which depends on the length of the laser pulse $\sigma$);
and (2) the energy spread of the electron beam distribution, $\sigma_E$. 
When the electron beam energy spread is small, the bandwidth
is dominated by the intrinsic single-electron scattering bandwidth. When the electron beam 
energy spread is high, electron beam's bandwidth dominates, thereby diminishing the effects
of the FM. This is illustrated in Fig.~\ref{fig:hw}.

A compensated narrow $n$th harmonic peaks near $E = 4\gamma^2 E_p n/(1+(1/2)a_0^2)$, 
where $E_p$ is the energy of the laser pulse.
This means that a desired radiation can be achieved either
by using the first harmonic from a backscattering off an electron beam with the 
energy $\gamma m_e c^2$, or the $n$th harmonic from a backscattering off an
electron beam with energy $\gamma m_e c^2/\sqrt{n}$. Such a
$1/\sqrt{n}$ reduction in the required electron beam energy would lead to 
substantial savings in the construction and shielding cost of the 
electron source. 

%
%
%
\begin{figure} 
\subfigure{
\includegraphics[width=3in]{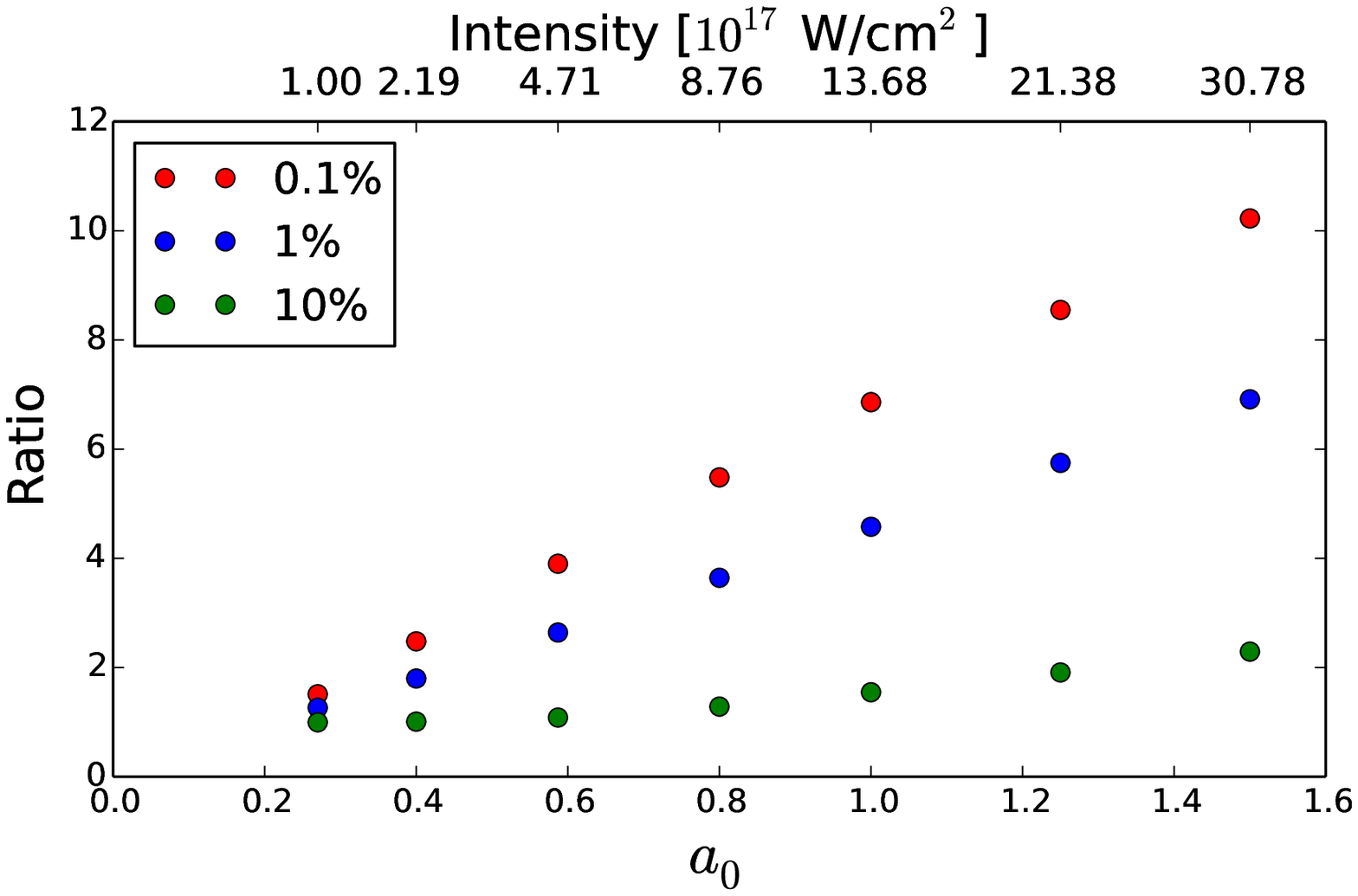} \label{fig:high_har1}}
\vskip-5pt
\subfigure{
\includegraphics[width=3in]{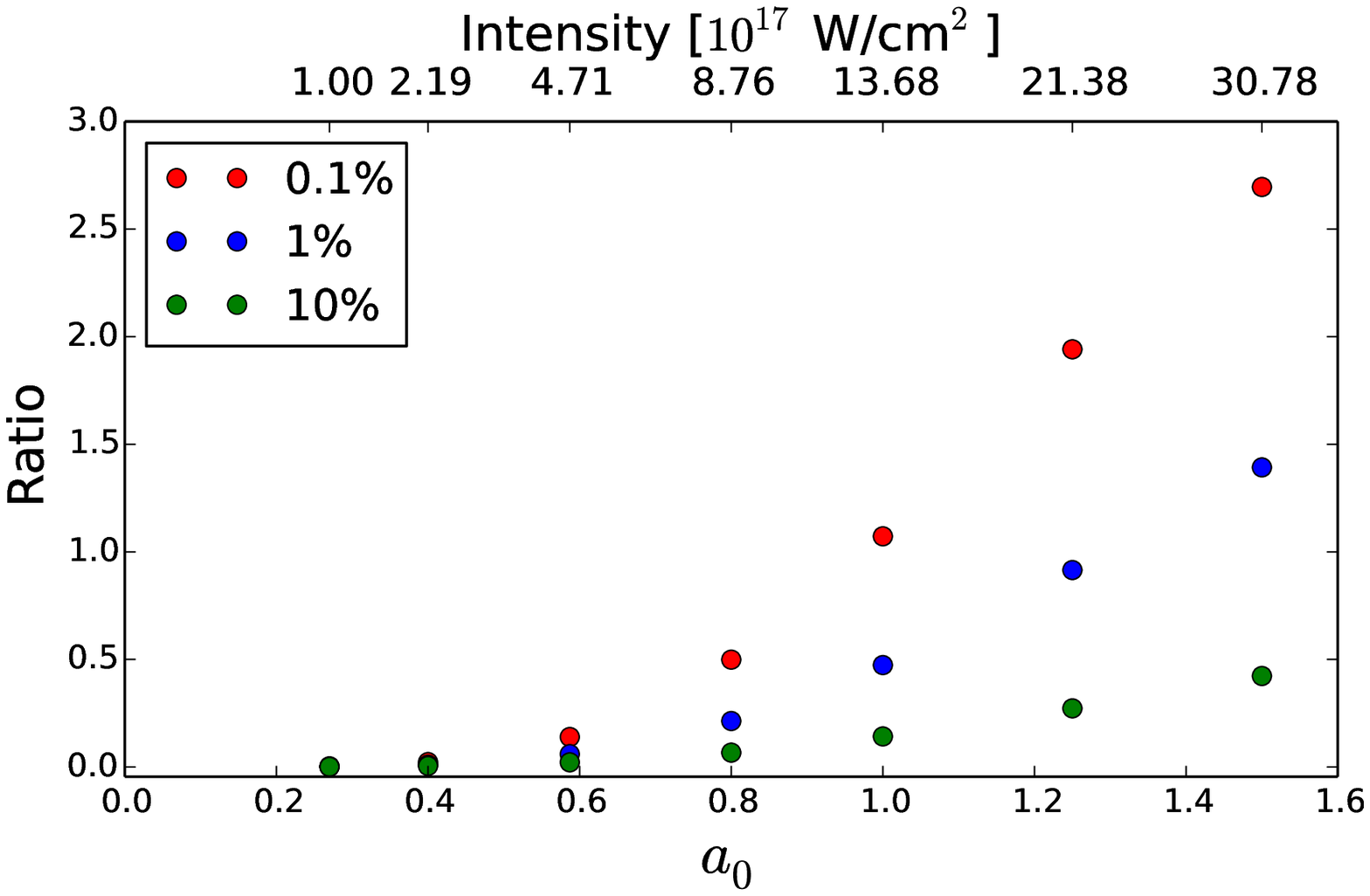} \label{fig:high_har3}}
\vskip-5pt
\subfigure{
\includegraphics[width=3in]{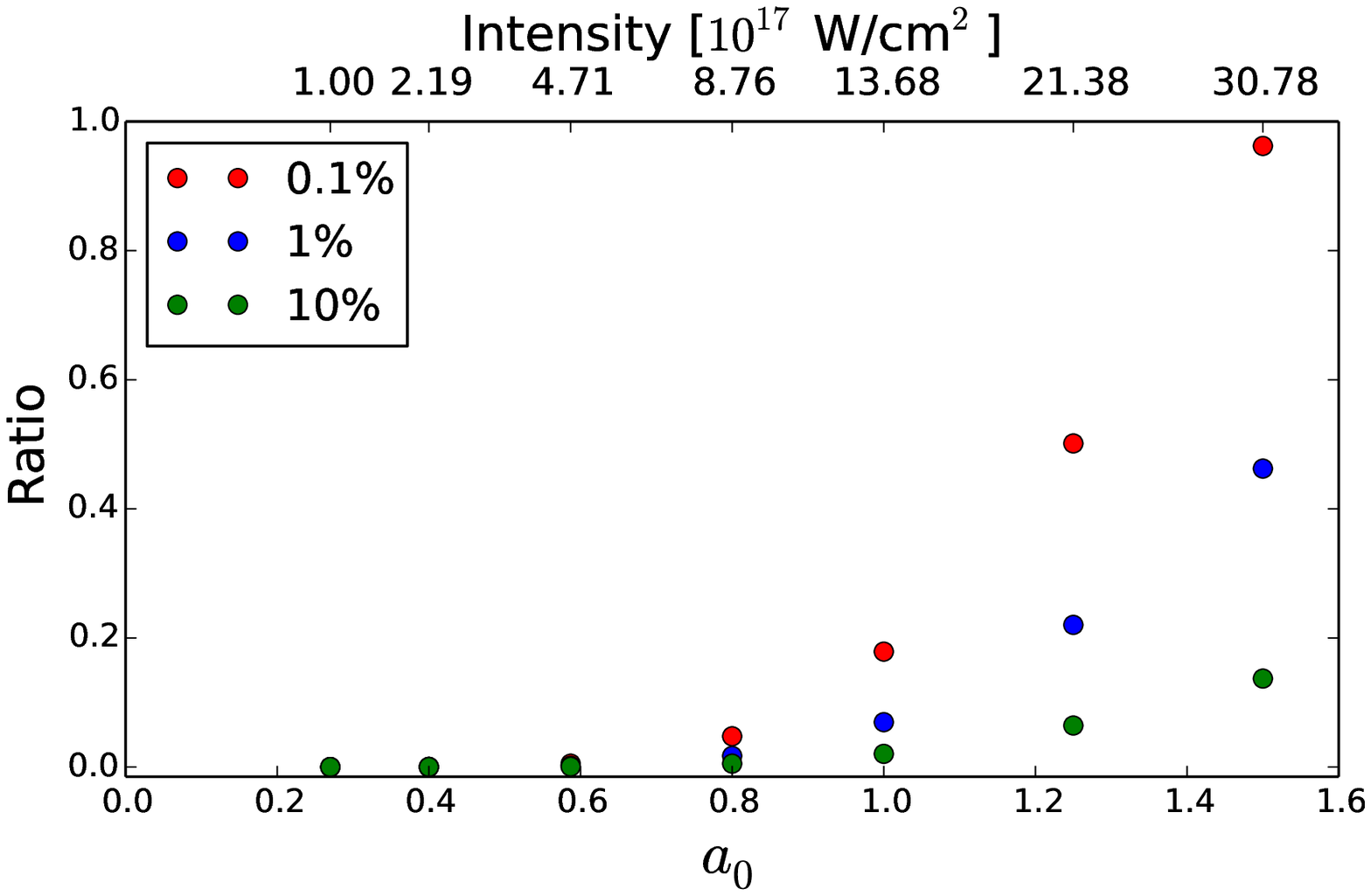} \label{fig:high_har5}}
\vskip-5pt
\caption{\small{Ratio of the maximum amplitudes as a function of the intensity
(and the amplitude of the normalized vector potential $a_0$) for a
Gaussian laser pulse with $\lambda = 1~\mu {\rm m}$ off a 
Gaussian electron beam with $Q = 100~{\rm pC}$, $E_0 = 163~{\rm MeV}$
and 0.1\% FWHM energy spread (red dots), 1\% (blue) and 10\% (green).
(a) FM first harmonic divided by the non-FM first harmonic.
(b) FM third harmonic divided by the non-FM first harmonic.
(c) FM fifth harmonic divided by the non-FM first harmonic.
The leftmost set of points ($a_0=0.27$ or $I = 10^{17}$ W/cm$^2$) 
corresponds to the current upper limit of operation of the laser-plasma
accelerators \cite{geddes2015}.}}
\label{fig:hw_har}
\end{figure}

Figure \ref{fig:hw_har} summarizes the importance of FM in Compton sources. FM
increases the yield in all harmonics to the point where for a high enough intensity,
the power in the higher harmonics exceeds that of the non-FM first harmonic.
For energy considered in this figure, 163 MeV, and the energy spread of 
0.1\% this occurs around $a_0=1$ (intensity of $1.4 \times 10^{18}$ W/cm$^2$) 
for the third FM harmonic and around $a_0=1.5$
($3.1 \times 10^{18}$ W/cm$^2$) for the fifth FM harmonic. These translate in
$\sqrt{3}$ and $\sqrt{5}$ reduction in electron beam energies, respectively.
As laser-plasma accelerator performance in the sub-1\% range in energy spread
becomes realistic \cite{geddes2015}, using the FM higher order harmonics
is poised to reap substantial benefits.

In this letter a novel calculation prescription is used to
determine the emission characteristics of the scattered radiation
in a Compton backscatter source. When compensated by laser
beam FM the radiation line heights and
widths may be accurately computed using a stationary phase argument
with crisp functional forms. The calculations accurately account for detuning
of the emitted radiation by beam energy spread.

Our calculations suggest the following main conclusion: by combining
harmonic generation and frequency modulation (chirping) of the incident
laser pulse it should be possible to generate significant fluxes of Compton
scattered radiation on the harmonic frequencies. The beam energy spread
in laser-plasma acceleration schemes are becoming good enough, 
nearing the 1\% level \cite{geddes2015}, that the energy flux 
density on the harmonics can even exceed that possible when an 
unmodulated incident laser pulse is used.

Using higher harmonics at
low laser intensities the photon yield is substantially
smaller than that in the first harmonic. The yield of the
higher harmonics can be improved---absolutely and
relatively compared to the first harmonic---by increasing
the laser intensity. However, increasing the laser intensity
leads to the ponderomotive broadening which 
erases any advantages thus incurred. Frequency modulation
and its perfect restoration of the narrowband emission, becomes crucial: it mitigates
the adverse effects of the ponderomotive broadening, allowing
one to continue increasing the photon yield.

C.~R.~acknowledges the support from the 
U.S.~Department of Energy, Science Undergraduate Laboratory Internship 
(SULI) program.



\bibliographystyle{elsarticle-num}




\end{document}